\documentclass[twocolumn,aps,prl,nofootinbib,superscriptaddress]{revtex4-2}
\usepackage{amsmath}
\usepackage{bm}
\usepackage{graphicx}
\usepackage{mathrsfs}
\usepackage{multirow}
\usepackage[normalem]{ulem}
\usepackage{graphicx}
\usepackage{footnote}
\usepackage{booktabs, ctable}
\usepackage{xcolor, framed}
\usepackage[colorlinks=true,
            linkcolor=red,
            citecolor=blue,
            urlcolor=black]{hyperref}

\begin{document}

\title{Femtoscopy as a New Probe of the Nuclear Equation of State}

\author{Xialei Jiang}
\email{xialeij@mails.ccnu.edu.cn}
\affiliation{Key Laboratory of Quark and Lepton Physics (MOE) and Institute of Particle Physics, Central China Normal University, Wuhan 430079, China}
\affiliation{GSI Helmholtzzentrum f\"{u}r Schwerionenforschung GmbH, Planckstrasse 1, 64291 Darmstadt, Germany}
 
\author{Jiaxing Zhao}
\email{jzhao@itp.uni-frankfurt.de}
\affiliation{Helmholtz Research Academy Hesse for FAIR (HFHF), GSI Helmholtz Center for Heavy Ion Physics, Campus Frankfurt, 60438 Frankfurt, Germany}
\affiliation{Institut f\"ur Theoretische Physik, Johann Wolfgang Goethe-Universit\"at,Max-von-Laue-Straße 1, D-60438 Frankfurt am Main, Germany}

\author{Yingjie Zhou}
\email{Z.Yingjie@gsi.de}
\affiliation{GSI Helmholtzzentrum f\"{u}r Schwerionenforschung GmbH, Planckstrasse 1, 64291 Darmstadt, Germany}

\author{Xiaofeng Luo}
\email{xfluo@ccnu.edu.cn}
\affiliation{Key Laboratory of Quark and Lepton Physics (MOE) and Institute of Particle Physics, Central China Normal University, Wuhan 430079, China}
 
\date{\today}

\begin{abstract}
Femtoscopic correlations are widely regarded as precision probes of hadronic interactions through vacuum final-state interactions after kinetic freeze-out. Here we demonstrate that, in baryon-rich heavy-ion collisions, the nuclear mean field generates an additional dynamical contribution to femtoscopic correlations during the transport evolution. Using the Parton--Hadron--Quantum--Molecular Dynamics (PHQMD) transport approach, we investigate proton--proton, proton--$\Lambda$, three-proton, and proton--proton--$\Lambda$ correlations in Au+Au collisions at $\sqrt{s_{\rm NN}}=3$, 4.5, 7.7, and 19.6 GeV. We find that the nuclear mean field produces a characteristic low-$k^*$ enhancement that is strongest at the lowest beam energies and gradually disappears with increasing collision energy. Furthermore, both the stiffness and the momentum dependence of the nuclear equation of state leave distinct signatures in the femtoscopic correlation functions, with higher-order correlations exhibiting substantially enhanced sensitivity compared with conventional two-particle observables. 
Our results demonstrate that femtoscopy extends beyond its traditional role as a tool for studying hadronic interactions and serve as a new class of microscopic observables for the nuclear equation of state, complementary to collective flow and subthreshold strangeness production, thereby opening a new avenue for exploring dense baryonic matter in low-energy heavy-ion collisions.
\end{abstract}

\maketitle

\emph{Introduction.--}
Determining the equation of state (EoS) of dense baryonic matter at supra-saturation densities remains one of the central challenges in modern nuclear physics. The high-density EoS governs the properties of neutron stars, core-collapse supernovae, and neutron-star mergers~\cite{Oertel:2016bki,Lattimer:2021emm,Burgio:2021vgk}, while also determining the pressure generated in compressed nuclear matter created in relativistic heavy-ion collisions. Despite its fundamental importance, the behavior of strongly interacting matter at high baryon density remains poorly constrained.

Significant progress has nevertheless been achieved over the past decades. At vanishing and small baryon chemical potential, first-principles lattice quantum chromodynamics (QCD) calculations predict a smooth crossover transition from hadronic matter to a deconfined quark-gluon plasma around $T_c\approx155~\rm MeV$~\cite{Aoki:2006we,Borsanyi:2010cj}. However, lattice QCD calculations at large baryon chemical potential remain hindered by the fermion sign problem, leaving the high-density region of the QCD phase diagram inaccessible from first principles. Considerable effort has therefore been devoted to constraining the high-density EoS, such as nuclear many-body theories~\cite{Gaitanos:2001hv,MALFLIET1988207}, effective models~\cite{Holt:2016pjb,Drischler:2021kxf,Stone:2006fn,Wang:2025uyl}, Bayesian inference~\cite{Pratt:2015zsa,Pang:2016vdc,OmanaKuttan:2022aml,Mohs:2024gyc}.

Heavy-ion collisions provide the only terrestrial laboratory capable of creating strongly interacting matter at several times the nuclear saturation density. During the compression stage, the pressure generated by the nuclear EoS is converted into experimentally measurable observables. Consequently, the high-density EoS has been extensively investigated through directed and elliptic flow~\cite{Molitoris:1986pp,Aichelin:1987ti,Kireyeu:2024hjo,Danielewicz:1998vz,Danielewicz:2000tr,Sorensen:2023zkk,Tarasovicova:2024isp}, stopping~\cite{Cozma:2024cwc,Sorensen:2023zkk}, cluster production~\cite{Sun:2018jhg,Zhou:2025zgn,Bratkovskaya:2025oys}, and subthreshold strangeness ($K^+$) production~\cite{Aichelin:1985rbt,Fuchs:2005zg,Hartnack:2005tr}. Despite considerable progress, extracting the EoS from heavy-ion collisions remains challenging because these observables are simultaneously influenced by the nuclear mean field, in-medium scattering, and nonequilibrium transport dynamics.

Femtoscopy has become a powerful tool for measuring the relative space-time structure of particle emission through quantum statistics and final-state interactions~\cite{Lisa:2005dd,NA49:2007fqa,Li:2008qm,Li:2022iil,Wiedemann:1996ig,Kisiel:2006is,Xu:2024dnd,Wang:2024bpl,Si:2025eou,Zhang:2017axr}. Besides determining the size and lifetime of the emitting source, recent measurements have demonstrated its capability to constrain hadron-hadron interactions, including hyperon-nucleon~\cite{STAR:2018uho,ALICE:2019buq,STAR:2005rpl,ALICE:2018ysd} and hyperon-hyperon~\cite{STAR:2014dcy,ALICE:2018ysd,ALICE:2022uso} interactions. Existing femtoscopic analyses generally treat the emission source as an input determined by the freeze-out geometry while attributing the correlation signal primarily to vacuum final-state interactions. The use of low-energy baryon femtoscopic correlations to constrain the density- and momentum-dependent nuclear EoS has remained largely unexplored.

At the beam energies of the RHIC Beam Energy Scan, HADES, CBM, NICA, and HIAF, baryons propagate in strong density- and momentum-dependent mean fields throughout the high-density stage. Different equations of state generate different pressure gradients, leading to distinct mean-field dynamics, trajectories, emission times, and relative space-time separations of emitted baryons. These differences are subsequently encoded in the femtoscopic correlations.
Unlike conventional EoS observables, which primarily probe the momentum-space response of compressed matter, femtoscopy directly measures its relative space-time structure, thereby providing a fundamentally different and complementary probe of the microscopic dynamics governed by the nuclear EoS.

In this Letter, we demonstrate that femtoscopic correlations provide a new probe of the nuclear equation of state in baryon-rich heavy-ion collisions. Using microscopic transport simulations with different nuclear mean-field parameterizations, we show that the stiffness and momentum dependence of the EoS leave characteristic signatures in two- and three-particle femtoscopic correlation functions. 

\emph{Framework.--}Our study is based on the  Parton-Hadron-Quantum-Molecular Dynamics (PHQMD)  microscopic transport approach  \cite{Aichelin:2019tnk,Glassel:2021rod,Kireyeu:2022qmv,Coci:2023daq}. PHQMD is a microscopic N-body transport model based on the QMD propagation of the baryonic degrees-of-freedom and the dynamical properties and interactions in- and out-of-equilibrium of hadronic and partonic degrees-of-freedom of the Parton-Hadron-String-Dynamics (PHSD) approach~\cite{Cassing:2009vt}.

The QMD equation-of-motions (EoM) for a N-body system are derived using the  Dirac-Frenkel-McLachlan  approach, which has  been developed in chemical physics and later applied to nuclear physics for QMD like models~\cite{Feldmeier:1989st,Aichelin:1991xy,Ono:1992uy,Hartnack:1997ez}. 
This approach is based on the variational formulation of the Schr\"odinger equation. With the assumption that the wave functions have a Gaussian form and that the width of the wave function is time independent, one obtains two equations-of-motion for the time evolution of the centroids of the Gaussian single particle Wigner density~\cite{Aichelin:1991xy}:
\begin{eqnarray}
\dot{{\bm r}}_{i}=\frac{\partial\langle H \rangle}{\partial {\bm p}_{i}} \qquad
\dot{{\bm p}}_{i}=-\frac{\partial \langle H \rangle}{\partial {\bm r}_{i}}.
\label{prop}
\end{eqnarray}
The Hamiltonian of the nucleus is the sum of the Hamiltonians of the nucleons, composed of kinetic and two-body potential energy, which has a strong interaction and a Coulomb part
\begin{eqnarray}
H = \sum_i H_i  = \sum_i  (T_i + \sum_{j\neq i}  V_{ij}).
\end{eqnarray}

\begin{figure}[!htb]
    \centering
    \includegraphics[width=0.45\textwidth]{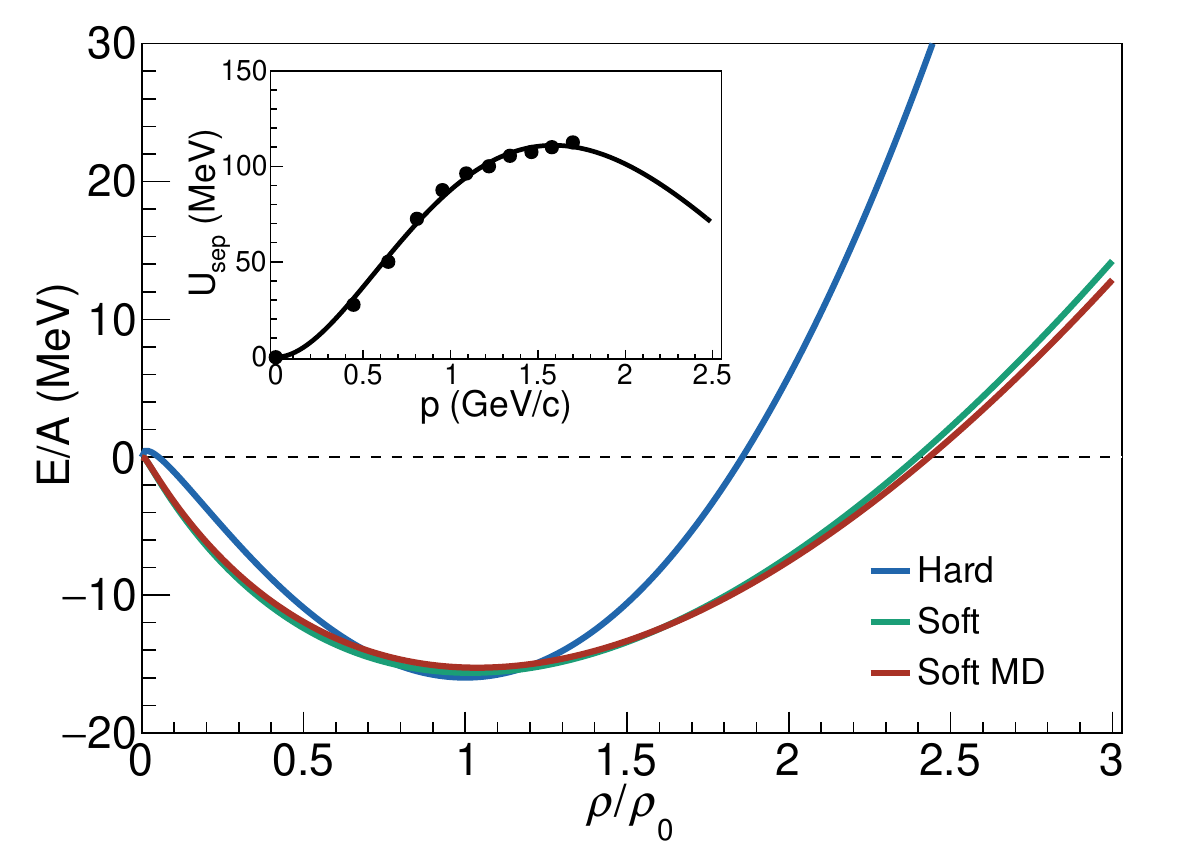} 
    \caption{The hard (blue line), soft (green line), and the soft momentum-dependent potential (red line) EoS at $T=0$. The inset panel is the momentum-dependent Schr\"odinger equivalent potential $U_{\rm sep}$.}
    \label{fig1}
\end{figure}

The expectation value of the Coulomb interaction, $\langle V_{\rm coul}\rangle$, can also be evaluated analytically. The expectation value of the Hamiltonian entering Eq.~(\ref{prop}) is then given by
\begin{eqnarray}
\langle H \rangle &=& \langle T \rangle + \langle V \rangle\nonumber\\
&=& \sum_i \big(\sqrt{{\bm p}_{i}^2+m^2}-m\big)+ \sum_{i} \langle V_{\rm Skyrme}({\bm r}_{i},t)  \nonumber\\ 
&+&V_{\rm Mom}({\bm r}_{i},{\bm p}_{i},t)+V_{\rm Coul}({\bm r}_{i},t)\rangle,
\end{eqnarray}
where $V_{\rm Skyrme}$ denotes the local Skyrme-type mean-field potential, $V_{\rm Mom}$ the momentum-dependent interaction, and $V_{\rm Coul}$ the Coulomb interaction. Their explicit expressions can be found in Refs.~\cite{Kireyeu:2024hjo,Zhou:2025zgn}. 
By averaging the two-body momentum-dependent interaction over the Fermi distribution of cold target nucleons, one obtains the Schr\"odinger-equivalent potential $U_{\rm sep}$, which has been extracted from elastic $pA$ scattering data~\cite{Cooper:2009zza,Cooper:1993nx}, as shown in the inset panel of Fig.~\ref{fig1}. For strange baryons, such as $\Lambda$ and $\Sigma$, the mean-field potential is taken to be $2/3$ of the nucleon potential, following the constituent quark model~\cite{Aichelin:2019tnk}.

In transport approaches, the nuclear interaction is commonly described by parametrized nucleon--nucleon or nuclear mean-field potentials, which correspond to different EoS of nuclear matter at zero temperature. To keep the parametrization as simple as possible, only a few free parameters are introduced. Two of them are fixed by the saturation properties of nuclear matter, requiring a binding energy per nucleon of $-16$ MeV at the saturation density $\rho_0$, while the remaining parameter determines the stiffness of the EoS and is conventionally characterized by the incompressibility modulus $K$,
\begin{eqnarray}
K= 9 \rho^2
\frac{{\rm \partial}^2(E/A(\rho))}{({\rm \partial}\rho)^2} \Big\vert_{\rho=\rho_0},
\end{eqnarray}
characterizing the curvature of the equation of state around the saturation density $\rho_0$. A smaller value of $K$ corresponds to a softer EoS with weaker resistance to compression, whereas a larger $K$ represents a hard (stiff) EoS. In this work, we employ the standard soft ($K=200$ MeV) and hard ($K=380$ MeV) Skyrme parameterizations~\cite{Molitoris:1986pp}, which approximately reproduce the incompressibility extracted from giant monopole resonances and high-density heavy-ion collision data, respectively~\cite{Mekjian:2011wut,Gustafsson:1984ka}. Besides, the soft momentum-dependent EoS is also adopted in this study as suggested by many previous studies~\cite{Hartnack:1997ez,Aichelin:1991xy,Kireyeu:2024hjo,Tarasovicova:2024isp,Zhou:2025zgn}. The comparison of different EoSs is shown in Fig.~\ref{fig1}.

\emph{Collision energy dependence.--}We employ the PHQMD code to simulate the particle evolution in Au+Au collisions at $\sqrt{s_{\rm NN}}=$3, 4.5, 7.7, and 19.6 GeV. The evolution is terminated at a given final time $t_f$. The corresponding kinematic selections are chosen to match those used in the RHIC Beam Energy Scan experiments. The two-particle correlation function is defined as the probability to find simultaneously two particles with momenta ${\bm p}_1$ and ${\bm p}_2$ divided by the product of the corresponding single particle probabilities. In experiments and simulations, the correlation is calculated via an effective way,
\begin{eqnarray}
C(k^*)= \mathcal{N}{N_{\rm same-events}(k^*)\over N_{\rm mixed-events}(k^*)},
\label{eq.corrmix}
\end{eqnarray}
where $N_{\rm same-events}(k^*)$ and $N_{\rm mixed-events}(k^*)$ are normalized relative momentum distributions of hadron pairs from the same and different (mixed) events, respectively. $k^*=|{\bm p}_1^*-{\bm p}_2^*|/2$ is the relative momentum in the center-of-mass frame of the pair. $\mathcal{N}$ is the normalization constant to make sure $C(k^*)\to 1$ at large $k^*$ using the interval $0.2<k^*<0.3$ GeV in the present study.

\begin{figure}[!htb]
    \centering
    \includegraphics[width=0.48\textwidth]{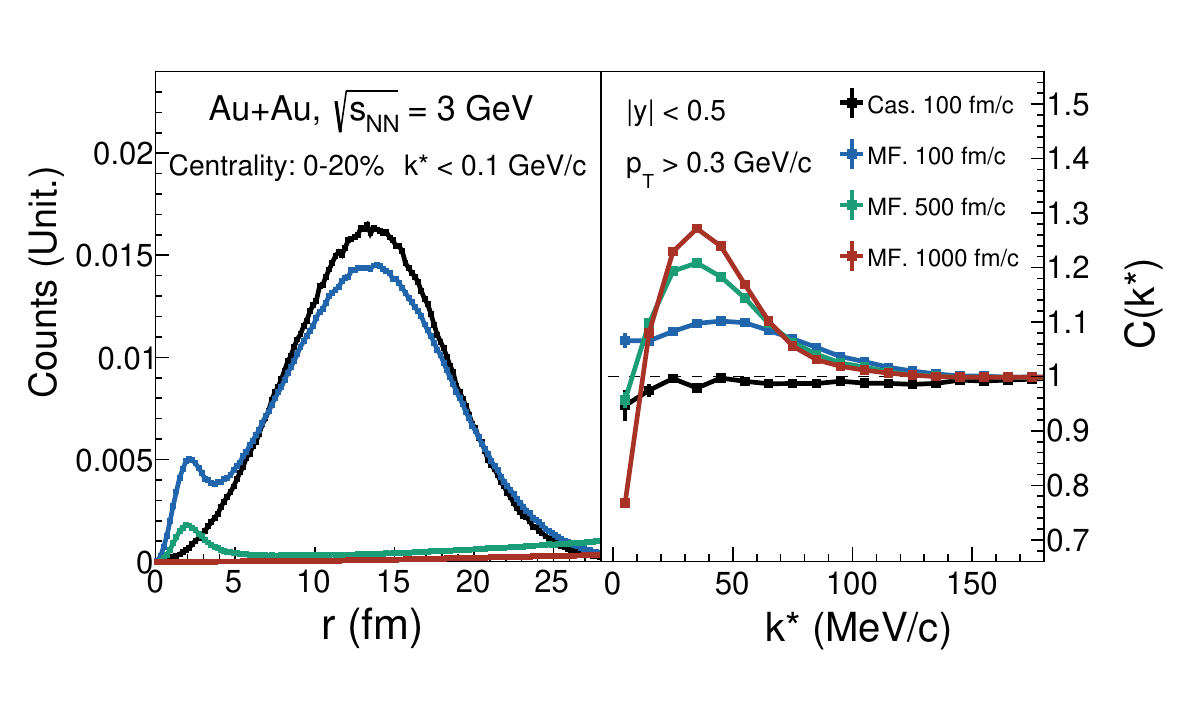} 
    \caption{The $p$-$p$ emission source (left) and correlation functions (right) with cascade (Cas. with black line: $t_f=100~\rm fm/c$) and mean-field (MF. with blue line: $t_f=100~\rm fm/c$; green line: $t_f=500~\rm fm/c$; red line: $t_f=1000~\rm fm/c$) simulations in 0-20\% Au+Au collisions at $\sqrt{s_{\rm NN}}=3~\rm GeV$.}
    \label{fig2}
\end{figure}
We first examine the proton--proton emission source, defined as the distribution of the relative distance between two protons in their center-of-mass frame. The results are presented in the left panel of Fig.~\ref{fig2}, where calculations with and without mean-field interactions (cascade) are compared. As shown in Fig.~\ref{fig2}, the inclusion of mean-field interactions with Hard EoS leads to a pronounced enhancement of the emission source at small relative distances $r$, resulting in a distinct peak structure that is absent in the cascade calculation.

The origin of this behavior is the collective nature of the nuclear mean field. Proton pairs with small relative momenta and distance experience nearly identical mean-field forces and therefore remain spatially correlated throughout the dynamical evolution. This generates a pronounced enhancement of the emission source at small relative distances, which is subsequently mapped onto the correlation function through the final-state interaction, producing the low-$k^*$ enhancement shown in the right panel of Fig.~\ref{fig2}. As the system continues to expand and dilute, the mean-field interaction gradually weakens, the short-range peak in the source disappears, and the correlation function correspondingly converges around $t_f=1000~\rm fm/c$. This value of $t_f$ is adopted in the following simulations.

We emphasize that the transport evolution is continued until the nuclear mean field has essentially vanished, by which time the dynamical correlations generated during the dense stage have converged and the emission source has expanded to a large spatial extent. Under these conditions, the short-range vacuum strong interaction is expected to provide only a limited correction to the mean-field-generated correlation. 

\begin{figure}[!htb]
    \centering
    \includegraphics[width=0.48\textwidth]{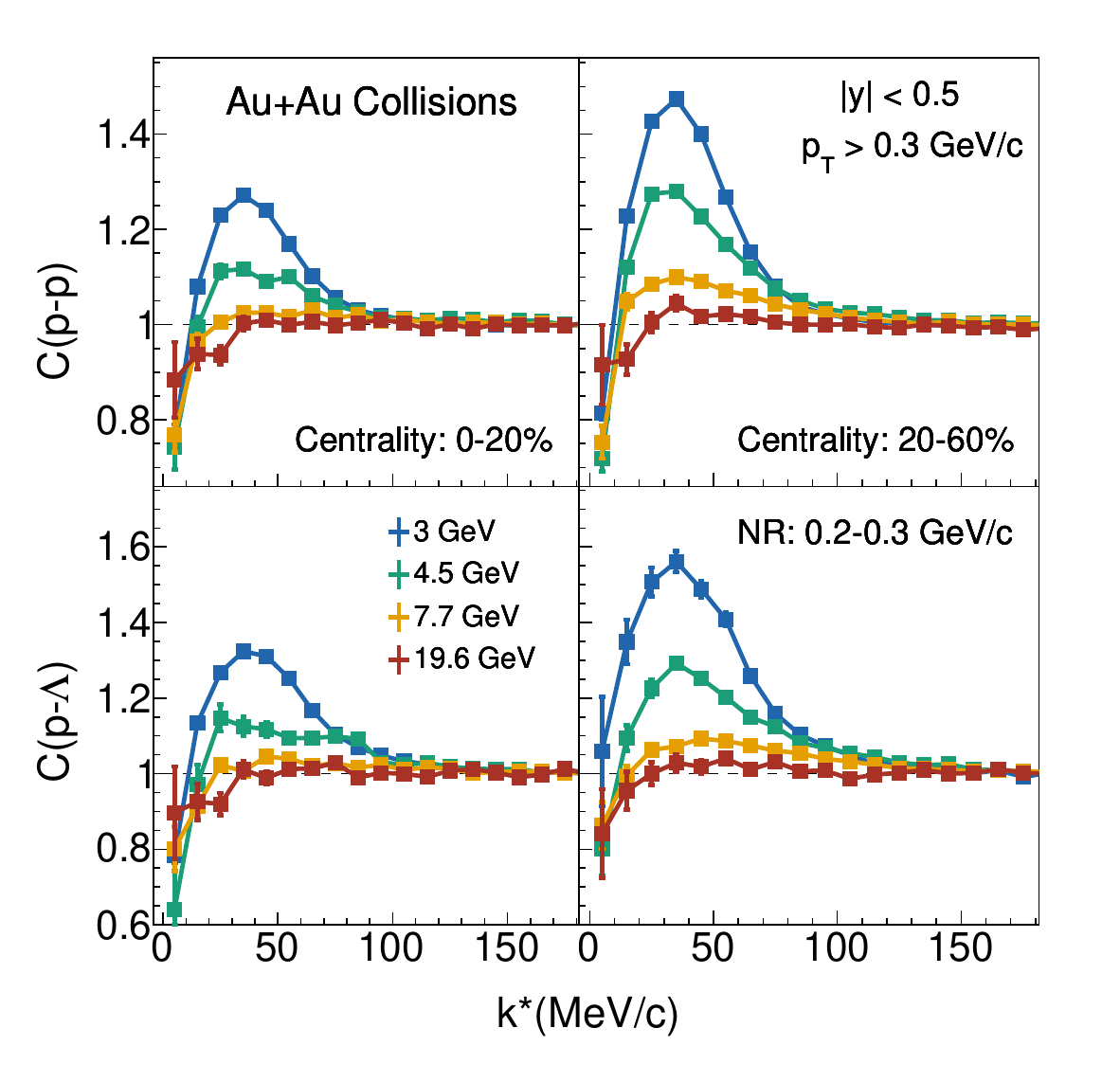} 
    \caption{Energy-dependent $p$-$p$ (upper panel) and $p$-$\Lambda$ (lower) correlations with Hard EoS in 0-20\% (left panel) and 20-60\% (right) Au+Au collisions at $\sqrt{s_{\rm NN}}=3,4.5,7.7,19.6~\rm GeV$.}
    \label{fig3}
\end{figure}
We now investigate the beam-energy dependence of the mean-field effect on femtoscopic correlations. Fig.~\ref{fig3} shows the proton--proton and proton--$\Lambda$ correlation functions at $\sqrt{s_{\rm NN}}=3$, 4.5, 7.7, and 19.6 GeV. A pronounced beam-energy dependence is observed. At $\sqrt{s_{\rm NN}}=3$ GeV, both correlation functions exhibit a strong low-$k^*$ enhancement, with peak values reaching $1.3$--$1.5$, depending on the particle species and collision centrality. The enhancement decreases rapidly with increasing beam energy, becomes modest at 7.7 GeV, and nearly disappears at 19.6 GeV.

The observed energy dependence reflects the diminishing role of the nuclear mean field at higher collision energies. At low beam energies, particles propagate for a longer time in the dense baryonic medium with relatively small momenta, allowing the mean field to generate substantial dynamical correlations between particle pairs. As the collision energy increases, the larger particle momenta and more rapid expansion shorten the time spent in the high-density region and reduce the influence of the mean field on the pair dynamics. Consequently, the mean-field-induced enhancement of the femtoscopic correlation function is progressively suppressed and becomes negligible at $\sqrt{s_{\rm NN}}=19.6$ GeV.

Compared to the central collisions, the correlation is considerably enhanced in the semi-central collisions, 20-60\%. This behavior is related to the different interplay between the nuclear mean field and the collective expansion. In semi-central collisions, the emitted baryons typically have lower average transverse momenta and undergo less thermal smearing~\cite{STAR:2024znc}, allowing the mean field to modify their relative motion more effectively. Consequently, stronger dynamical correlations are generated during the transport evolution. A similar mechanism explains the more pronounced correlation observed in the $p$-$\Lambda$ system, where the lower average momentum of hyperons enables them to remain under the influence of the mean field for a longer time, thereby enhancing the dynamical correlation~\cite{STAR:2024znc}.

\begin{figure}[!htb]
    \centering
    \includegraphics[width=0.48\textwidth]{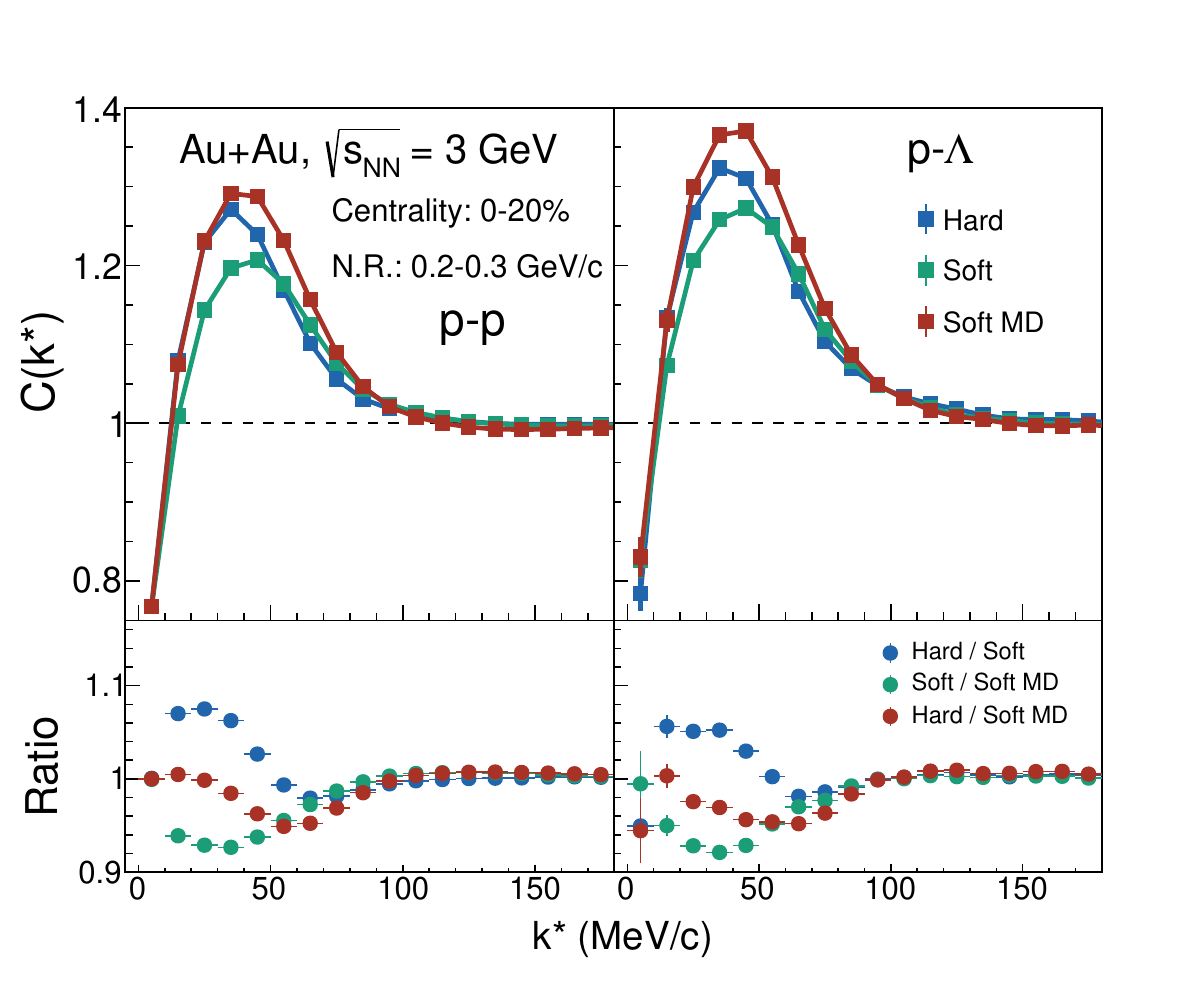} \\
    \includegraphics[width=0.48\textwidth]{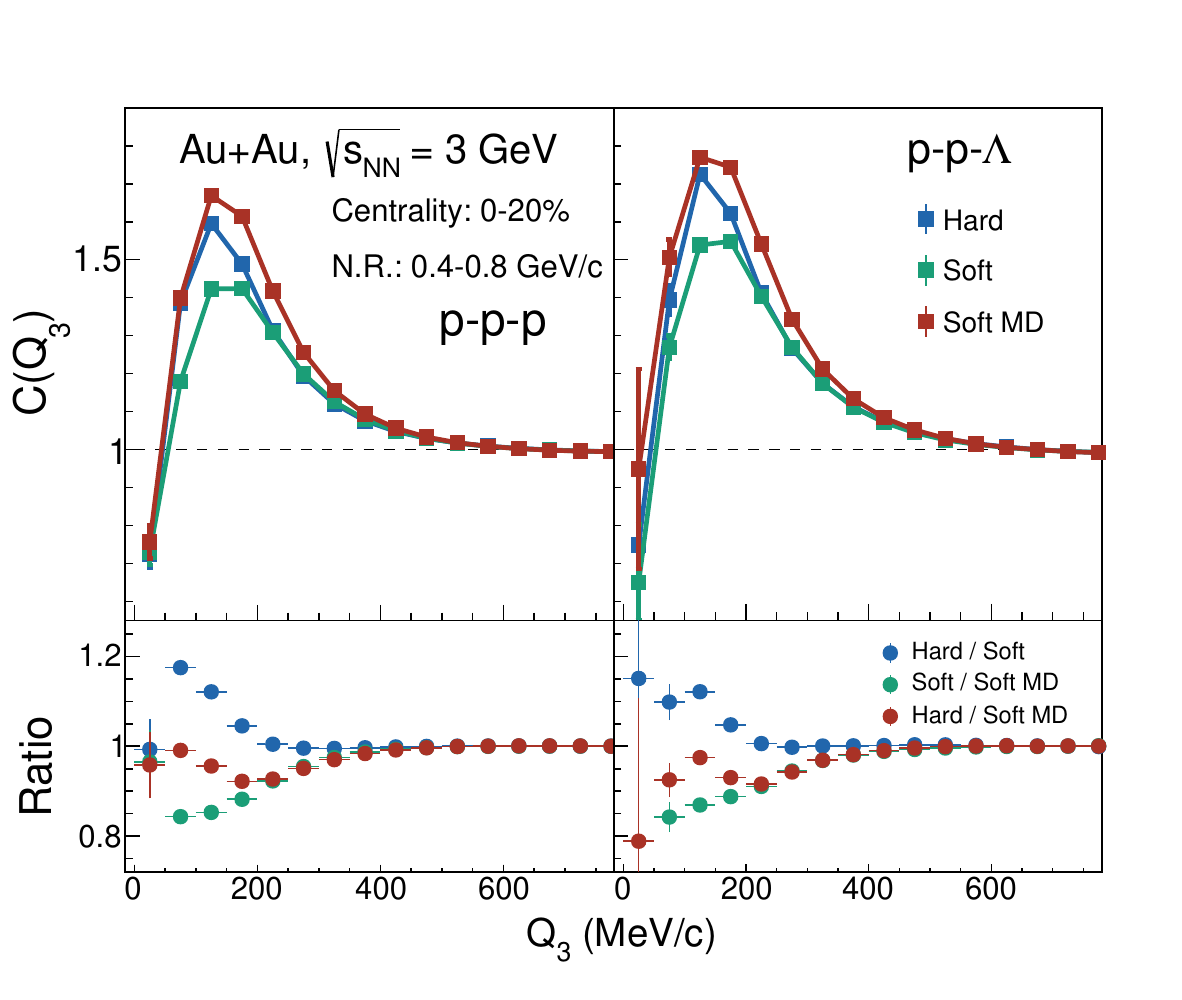} 
    \caption{The two-particle (upper panel; left: $p$-$p$, right: $p$-$\Lambda$) and  three-particle (lower panel; left: $p$-$p$-$p$, right: $p$-$p$-$\Lambda$) correlation functions with different EoSs in 0-20\% Au+Au collisions at $\sqrt{s_{\rm NN}}=3~\rm GeV$. In each panel, the lower subpanel shows the ratios of the correlation functions for the Hard to Soft, Soft to Soft momentum-dependent (Soft MD), and Hard to Soft MD EoS.}
    \label{fig4}
\end{figure}

\emph{Probe the equation of state.--}
Having established that the nuclear mean field generates characteristic two-particle correlations, we now investigate whether higher-order femtoscopy can further enhance the sensitivity to the nuclear equation of state. Three-particle femtoscopy probes three-body phase-space correlations and is therefore expected to amplify collective effects generated during the dynamical evolution.

The three-particle correlation function is defined as
\begin{eqnarray}
C(Q_3)= \mathcal{N}{N_{\rm same-events}(Q_3)\over N_{\rm mixed-events}(Q_3)},
\label{eq.corrmix3}
\end{eqnarray}
where $N_{\rm same-events}(Q_3)$ and $N_{\rm mixed-events}(Q_3)$ are normalized relative momentum distributions of hadron triplets from the same and mixed events, respectively. The Lorentz-invariant variable $Q_3$ is defined as,
$Q_3=\sqrt{-q_{12}^2-q_{23}^2-q_{31}^2}$ with the relative four-momentum $q_{ij}^\mu=(p_i-p_j)^\mu-(p_i-p_j)^\nu P_{ij,\nu}/(P_{ij}^2) P_{ij}^\mu $. $P_{ij}^\mu=p_i^\mu+p_j^\mu$ is total momentum of particles $i$ and $j$~\cite{ALICE:2013uhj,ALICE:2022boj}.

Figures~\ref{fig4} present the two-particle ($p$-$p$, $p$-$\Lambda$) and the three-particle ($p$-$p$-$p$, $p$-$p$-$
\Lambda$) correlation functions in central Au+Au collisions at $\sqrt{s_{\rm NN}}=3$ GeV for the hard, soft, and soft momentum-dependent equations of state. A clear EoS dependence is observed in all four correlation functions. The soft momentum-dependent EoS (red lines) produces the strongest correlation, followed by the hard EoS, whereas the soft EoS consistently yields the weakest enhancement. 
This ordering demonstrates that, in addition to the incompressibility of nuclear matter, the momentum dependence of the nuclear mean field plays a crucial role in generating femtoscopic correlations. Compared with the soft EoS, the larger incompressibility of the hard EoS strengthens the collective expansion and consequently enhances the dynamical correlations among emitted particles. The soft momentum-dependent EoS further introduces a velocity-dependent contribution, $\partial V_{\rm Mom}/\partial {\bm p}$, into Hamilton's equations of motion. For particle pairs with small relative momenta and separations, the two particles experience nearly identical momentum-dependent mean-field evolution and therefore follow similar space-time trajectories over an extended period of the reaction. This correlated propagation preserves the dynamical correlation between the particles and results in the strongest correlations.

The lower panels show the ratios between different EoS calculations, highlighting the EoS sensitivity more clearly. Around the correlation peak, the two-particle correlations exhibit variations of approximately 5-10\%, whereas the three-particle correlations show substantially larger differences of up to $\sim$20\%. A similar ordering is found for $p$-$p$-$\Lambda$ femtoscopy, indicating that the effect is a generic consequence of the mean-field dynamics rather than a feature of a particular particle species.

These results demonstrate that femtoscopic correlations retain clear sensitivity to both the stiffness and the momentum dependence of the nuclear equation of state. In particular, higher-order femtoscopy amplifies the many-body dynamical correlations generated by the nuclear mean field, making it a powerful and complementary probe of dense baryonic matter in low-energy heavy-ion collisions.

\emph{Summary.--} 
In this Letter, we have investigated proton--proton, proton-$\Lambda$, three-proton, and proton-proton-$\Lambda$ correlations in Au+Au collisions at $\sqrt{s_{\rm NN}}=3$, 4.5, 7.7, and 19.6 GeV within the PHQMD transport approach. By systematically comparing calculations with different nuclear mean-field interactions and equations of state, we arrive at the following conclusions.

(i) We demonstrate that the nuclear mean field generates a previously overlooked dynamical contribution to femtoscopic correlations, in addition to the conventional vacuum final-state interaction in low-energy heavy-ion collisions. This mean-field-induced correlation originates from the collective propagation of baryons in the dense medium and produces a characteristic low-$k^*$ enhancement of the correlation function. Its magnitude increases rapidly with decreasing beam energy and becomes particularly pronounced in the baryon-rich regime. 

(ii) The mean-field-induced correlations exhibit pronounced sensitivity to both the stiffness and the momentum dependence of the nuclear equation of state. While conventional two-particle femtoscopy already discriminates between different EoS parameterizations, higher-order femtoscopic correlations further amplify the EoS dependence by probing genuine many-body dynamical correlations generated during the transport evolution. Our results therefore demonstrate femtoscopy, and in particular higher-order femtoscopy, as a new class of microscopic observables for the nuclear equation of state. Complementary to traditional probes such as collective flow and subthreshold particle production, femtoscopic correlations provide direct access to the space-time response of dense baryonic matter, opening a new avenue for exploring the equation of state in baryon-rich heavy-ion collisions.

\textbf{Acknowledgment.--} J. Zhao acknowledges the support by the Deutsche Forschungsgemeinschaft (DFG) through the grant CRC-TR 211 "Strong-interaction matter under extreme conditions" (Project number 315477589 - TRR 211). X. Luo is supported in part by the National Natural Science Foundation of China under Grant No. 12525509, No. 12447102, and the National Key Research and Development Program of China under contract No. 2022YFA1604900, and the Fundamental Research Funds for the Central Universities (XJ2026000302). The computational resources utilized for this work were provided by the Center for Scientific Computing (CSC) at Goethe University Frankfurt.

\bibliographystyle{apsrev4-2}
\bibliography{refs}

\end{document}